# Bound State in the Continuum in Slab Waveguide Enables Low-Threshold Quantum-Dot Lasing

Mengfei Wu[1,2], Lu Ding[1], Randy P. Sabatini[2], Laxmi Kishore Sagar[2], Golam Bappi[2], Ramón Paniagua-Domínguez[1], Edward H. Sargent[2,*], Arseniy I. Kuznetsov[1,*]

[1] Institute of Materials Research and Engineering, A*STAR (Agency for Science, Technology and Research), Singapore 138634, Singapore

[2] Department of Electrical and Computer Engineering, University of Toronto, Toronto, Ontario M5S 3G4, Canada

**ABSTRACT:** Colloidal quantum dots (CQDs) are a promising gain material for solution-processed, wavelength-tunable lasers, with potential application in displays, communications, and biomedical devices. In this work, we combine a CQD film with an array of titanium dioxide ($TiO_2$) nanoantennas to achieve lasing via bound states in the continuum (BICs), which are symmetry-protected cavity modes with giant quality factors. Here, the BICs arise from slab waveguide modes in the planar film, coupled to the periodic nanoantenna array. We engineer the thickness of the CQD film and size of the nanoantennas to achieve a BIC with good spatial and spectral overlap with the CQDs, based on a $2^{nd}$-order TE-polarized waveguide mode. We obtain room-temperature lasing with a low threshold of approximately 11 kW/cm$^2$ (peak intensity) under 5 ns-pulsed optical excitation. This work sheds light on the optical modes in solution-processed, distributed-feedback lasers, and highlights BICs as effective, versatile, surface-emitting lasing modes.

Solution-processed lasers, such as those made using colloidal quantum dots (CQDs),[1–7] colloidal quantum wells,[8–11] and perovskites,[12–14] possess wide spectral tunability and a compact, flexible form factor, and are of interest for display, communications, sensing, and biomedical applications. Significant material innovations have led to continuous-wave, solution-processed lasers at or close to room temperature, with a threshold pump intensity ranging from ~ 100 W/cm$^2$ to ~ 10 kW/cm$^2$.[3,8,13] Breakthroughs derived from innovations in optical design have also been reported.[10,15]

In general, a low-threshold laser requires not only an optical cavity with minimal absorption or radiation loss, and hence a high quality (Q) factor, but also good spatial and spectral overlap between the cavity mode and gain material to ensure efficient light-matter interaction. In previous solution-processed lasers, distributed feedback (DFB)[2–4,12,13] and Fabry-Pérot[1,8,9] cavities have been employed extensively. A Fabry-Pérot cavity can possess a high Q factor and can position the gain material at the field maximum; however, fabrication of a high-Q device is challenging because of the large number of layers required and the need for precisely controlled layer thicknesses. A DFB cavity, on the other hand, typically consists of a grating made of a dielectric, coated with a solid film of gain material. The spatial mode-gain overlap extends laterally across the entire device, and the in-plane reflections (as a result of the periodic variation in refractive index) can produce a high Q factor when the number of periods is large. Commercially available DFB lasers made of III-V semiconductors rely on the fundamental DFB mode, which involves a phase shift of $\pi$ per period, to achieve edge lasing. In contrast, reported solution-processed lasers have largely relied on the 2$^{nd}$-order DFB mode, which contains a phase shift of 2π per period. While optical feedback is in plane, a 2nd-order DFB can couple light out of plane via diffraction to achieve lasing in the normal direction.

Upon closer examination, the 2nd-order DFB involves two distinct band-edge modes with different field distributions, representing the two ways a standing wave could be formed from two counter-propagating waves in the grating. This is illustrated for transverse-electric (TE)-polarized modes in Figures 1a and 1b, which plot the y-component of the electric field of the standing wave along a grating. The field of the mode in Figure 1a has the same symmetry as that of a plane wave at normal incidence (e.g. with respect to a y-z plane cutting in the middle of a unit cell); the mode can thus couple to the far field by diffraction. In contrast, the mode in Figure 1b has a different symmetry than the plane wave and therefore cannot couple to the far field in the normal direction. With a lower radiation loss, the mode in Figure 1b should have a higher Q factor. In fact, its Q factor could be infinite in theory for an ideal system without any absorption, scattering, or diffractive losses. Such a mode is known as a symmetry-protected bound state in the continuum (BIC).[16–20]

Taking advantage of the impressive Q factors, BICs have been utilized to achieve lasing in III-V semiconductors,[21–26] colloidal quantum wells,[11] and perovskites.[14,27] Early reports on CQD lasers,[2–6] however, did not differentiate between a BIC and a diffractively coupled band-edge mode. In the present work, we show that lasing occurs through symmetry-protected BICs, by experimentally determining the dispersion diagram (i.e. the photonic band structure). In addition, we engineer the CQD film thickness so that the grating-CQD slab supports not only 1st-order but also 2nd-order waveguide modes (Figure 1c). We show that the BIC arising from a 2nd-order waveguide mode gives a lower lasing threshold, thanks to better spatial mode-gain overlap. This leads to room-temperature lasing under 5 ns-pulsed optical excitation with a threshold peak intensity of approximately 11 kW/cm$^2$.

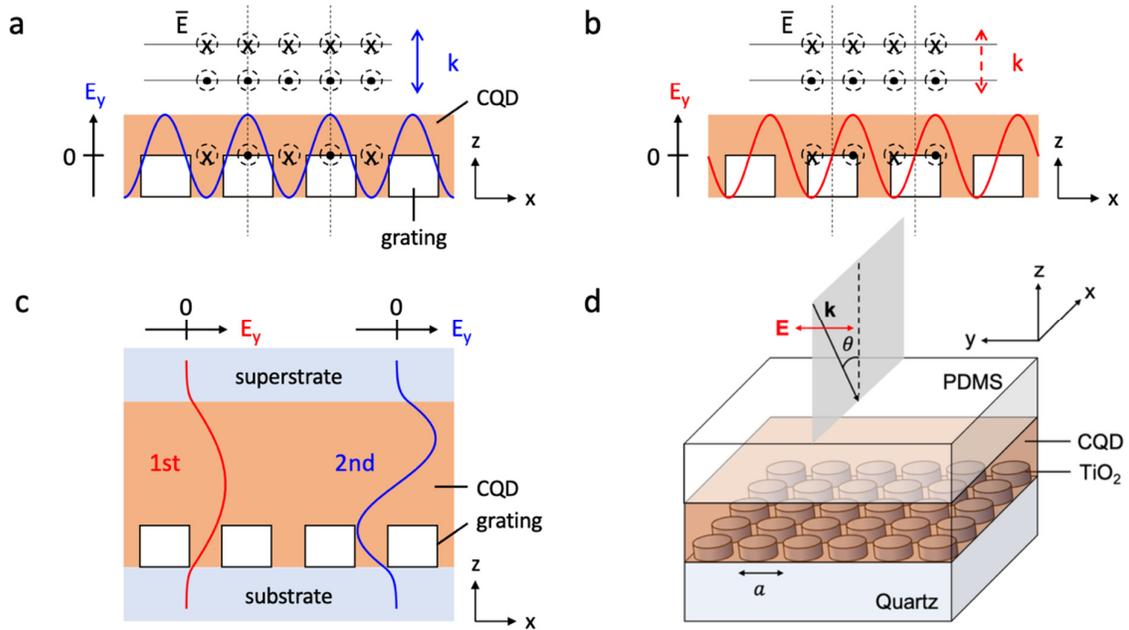

**Figure 1.** (a, b) Schematic showing two distinct 2$^{nd}$-order DFB modes with transverse-electric (TE) polarization: (a) diffraction-coupled band-edge mode, (b) bound state in the continuum (BIC). The y-component of the electric field ($E_y$) is plotted for the standing wave along the x direction of the grating. The direction of the electric field ($\bar{E}$) is indicated for the standing wave as well as the plane wave incident on the structure in the normal direction. The field of a BIC has a different symmetry compared to that of the plane wave, so a BIC is decoupled from the far field and has zero radiation loss. CQD: colloidal quantum dot. (c) Sketch of $E_y$ along the transverse (z) direction for the 1$^{st}$- and 2$^{nd}$-order slab waveguide modes supported in the grating coated with a relatively thick layer of CQDs. (d) Device structure of the CQD laser in this work. The TiO$_2$ cylinders are 120 nm in height, and the period of the square array is tuned with the cylinder diameter, keeping the gap between the cylinders fixed at 40 nm. The CQD film is about 300 nm in thickness. Optical measurements are taken in the x-z plane with TE polarization.

In the CQD laser reported herein (Figure 1d), the grating is a 2D square array of titanium dioxide ($TiO_2$) cylindrical nanoantennas, with a height of 120 nm and a gap of 40 nm between cylinders in both x and y directions. Keeping the gap distance constant, we adjust the period of the grating by changing the diameter of the cylinders. Here, we use $TiO_2$ instead of the typical silicon dioxide ($SiO_2$) in earlier reports of DFB lasers since, with the relatively high refractive index of $TiO_2$ ($n \sim 2.5$), Mie resonances can be excited efficiently in the cylinders, which enhances the optical intensity in the near field, thus allowing stronger light-matter interaction.[11,28–30] The CQDs spin-coated on the $TiO_2$ grating have been engineered to facilitate population inversion and hence lower the optical gain threshold.[3] They consist of a cadmium selenide (CdSe) core and an asymmetric cadmium sulfide (CdS) shell capped with chloride ligands (see Methods in the Supporting Information for details). The refractive index of these CQDs is ~ 1.9 at visible wavelengths, as measured using spectroscopic ellipsometry. Upon spin coating, the CQD film is approximately 300 nm thick on a plain (grating-free) area, as measured using a profilometer. The sample is then capped with a layer of polydimethylsiloxane (PDMS, ~ 1 mm thick), whose refractive index is similar to that of the quartz substrate ($n \sim 1.5$).

Given the thickness and refractive indices, the combined CQD-$TiO_2$ slab is able to support 1st- and 2nd-order waveguide modes. For both modes, at a given frequency, the grating periodicity can be tuned so that the longitudinal (in-plane) component of the wavevector gives a phase delay of $2\pi$ per period of the grating. Under these conditions counter-propagating waves form a standing wave, resulting in either a band-edge mode radiating in the normal direction via diffraction or a symmetry-protected BIC with no (or suppressed) radiation in the normal direction. The two modes have a slight energy difference owing to the different localizations of the field in the grating, giving rise to a bandgap at the $\Gamma$ point in the dispersion diagram.

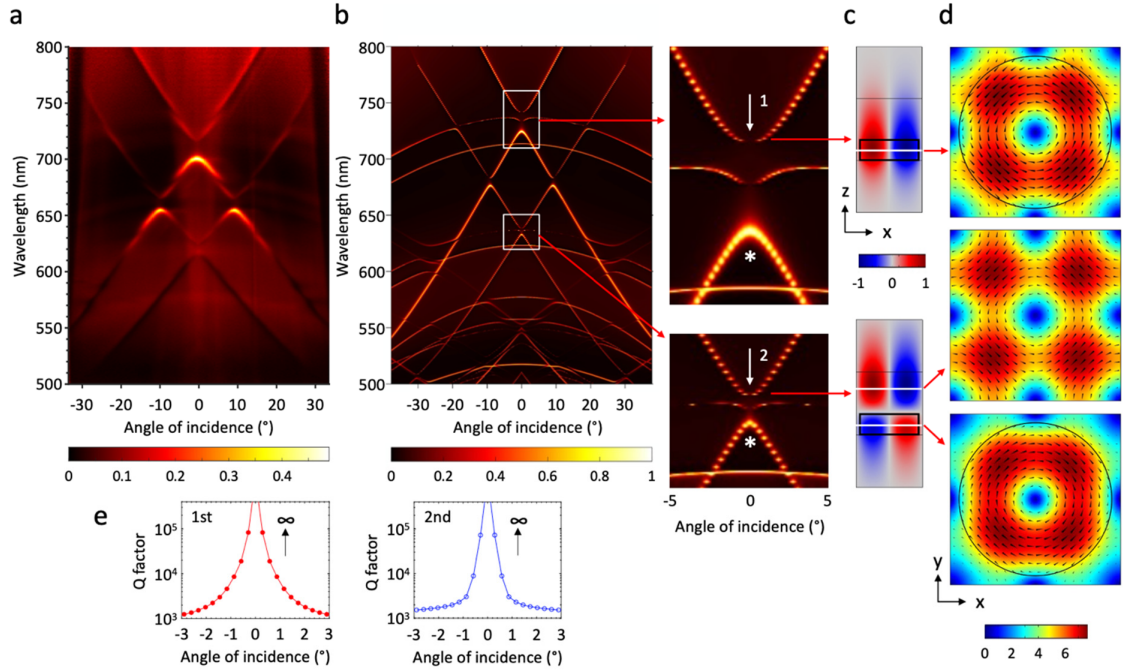

**Figure 2.** (a) Angle-resolved reflectance spectrum measured for a CQD-coated $TiO_2$ array with a cylinder diameter of 340 nm, under TE-polarized incident light. (b) Corresponding reflectance spectrum obtained by numerical simulations, assuming no absorption, flat surfaces, and infinite array size. The regions of interest are boxed and magnified for clarity, showing bands arising from the 1st- and 2nd-order slab waveguide modes in the upper and lower panels, respectively. Arrows point to BICs, while asterisks (*) indicate diffraction-coupled band-edge modes. (c) The y-component of the electric field ($E_y$) plotted in the x-z plane cut along the middle of a unit cell for the two BICs. (d) Amplitude (colour map) and polarization vectors (arrows) of the electric field plotted in the x-y plane cut along the white lines indicated in (c). The $TiO_2$ cylinders are outlined in black in (c) and (d). (e) Quality (Q) factor as a function of the incident angle in the vicinity of the 1st- (left) and 2nd-order (right) BICs, derived from (b). The Q factors in practice are lower due to absorption, sample imperfection, and a finite sample size.

To determine experimentally the photonic band structure of the sample, we measure the reflectance spectrum resolved in both wavelength and angle of incidence (or reflection), by projecting the back focal plane image through the slit of a spectrometer (see Methods in the Supporting Information for details). Figure 2a shows the reflectance spectrum of a CQD-coated $TiO_2$ array with a cylinder diameter of 340 nm and a period of 380 nm, when the incident light is TE-polarized. We observe bands arising from diffractive coupling of the 1st- and 2nd-order waveguide modes, as well as vanishing reflectance at normal incidence near the bandgaps, indicating the presence of BICs. To understand the optical modes better, we run a numerical simulation to obtain the more detailed dispersion diagram in Figure 2b. Here, we see salient features of several BICs at the Γ point (0°), where reflectance is inhibited since incident plane waves cannot couple to the BICs due to symmetry incompatibility.

The arrows in Figure 2b indicate the two BICs of interest, and eigenmode calculations give their mode profiles (Figures 2c and 2d for the side and top views of a unit cell, respectively). As expected, the BIC at $\lambda \sim 740$ nm arises from a 1st-order TE-polarized waveguide mode, while the BIC at $\lambda \sim 630$ nm originates from a 2nd-order waveguide mode. The in-plane polarization vectors (Figure 2d) reveal a circulating electric field, which is analogous to the field of an out-of-plane magnetic dipole. Thus, the BICs we observe here are similar to the previously reported BIC based on an array of magnetic dipoles oscillating in phase in the normal direction.[11,31,32]

Lastly, we derive the theoretical Q factors of the BICs from the dispersion diagram. Figure 2e shows that for each BIC, the Q factor increases towards infinity as the angle of incidence approaches zero. The simulation assumes that the extinction coefficients (κ) of all materials in the system are zero, all surfaces are smooth, and the periodic array is infinite in size. In reality, the CQDs have finite κ at visible wavelengths, the surface of the film is slightly undulating on top of the grating, and the array is 50 × 50 μm in size. As a result, the Q factors are finite for the BICs observed experimentally, also known as quasi-BICs.[16,17,33] Unfortunately, due to the limited resolution of our spectrometer (~ 2 nm; see Methods in the Supporting Information for details), we are unable to measure the high Q factors from the sample directly.

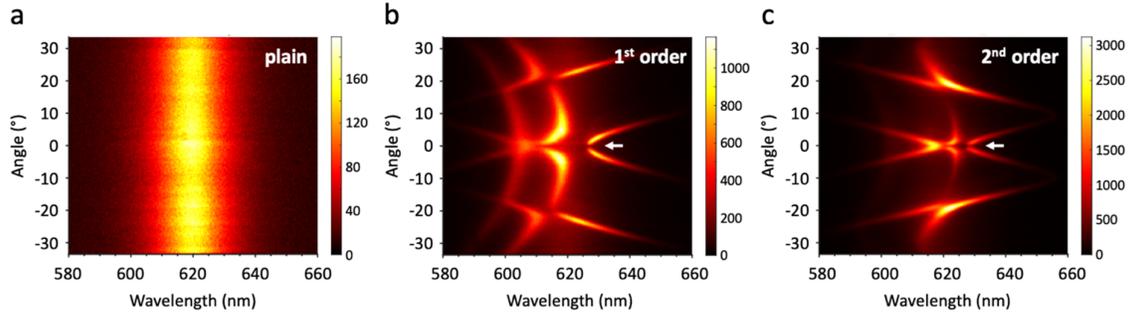

**Figure 3.** (a, b, c) Angle-resolved PL spectra of CQDs measured on (a) plain area of the substrate, (b, c) a $TiO_2$ array with a cylinder diameter of (b) 280 nm and (c) 340 nm. The sample is excited with a continuous-wave laser at $\lambda = 488$ nm, and the PL is collected through a polarizer for TE-polarized light only. (b) and (c) support BICs originating from the 1$^{st}$- and 2$^{nd}$-order slab waveguide modes, respectively, marked by arrows.

With the goal of achieving lasing, we first characterize the coupling of the CQD photoluminescence (PL) to the BICs under the excitation of a continuous-wave (CW) laser at $\lambda = 488$ nm. The angle-resolved PL spectrum of a plain CQD film (Figure 3a) shows an angle-independent PL centered at $\lambda = 620$ nm with a full width at half maximum (FWHM) of 20 nm; see Figure S1 in the Supporting Information for a PL spectrum integrated across angles. We then seek to match the 1$^{st}$- and 2$^{nd}$- order BICs spectrally to the CQD PL by varying the diameter of the $TiO_2$ cylinders keeping the gap between cylinders fixed at 40nm. The larger the cylinder diameter (i.e. the larger the grating period), the more red-shifted the optical modes, as the required in-plane propagation constant becomes lower for a given phase delay per period. For the cylinder array presented in Figure 2, the 2$^{nd}$-order BIC is in the desired spectral range ($\lambda \sim 630$ nm). To match the gain peak with the 1st-order BIC, however, the resonances must be blue-shifted. We find that when the cylinder diameter is reduced to 280 nm (period to 320 nm), the resonant wavelength of the 1st-order BIC is shifted to $\lambda \sim 630$ nm. Figures 3b and 3c contain the TE-polarized PL spectra of the CQD-coated arrays, with a cylinder diameter of 280 nm and 340 nm, respectively, showing efficient coupling of the CQD PL to the various waveguide modes. In particular, we observe clear symmetry-protected BICs with a radiation node in the normal direction, namely the 1st- and 2nd-order BICs in Figures 3b and 3c, respectively. The light emitted by the CQDs that is coupled to the BICs cannot radiate into the far field due to a mismatch in the field symmetry and is thus trapped in the cavity, providing efficient optical feedback for the CQDs. Furthermore, comparing Figures 3c and 2a, we notice that the PL map reflects the photonic bands more clearly than the reflectance map, potentially because of the additional enhancement in CQD emission by the Purcell effect.

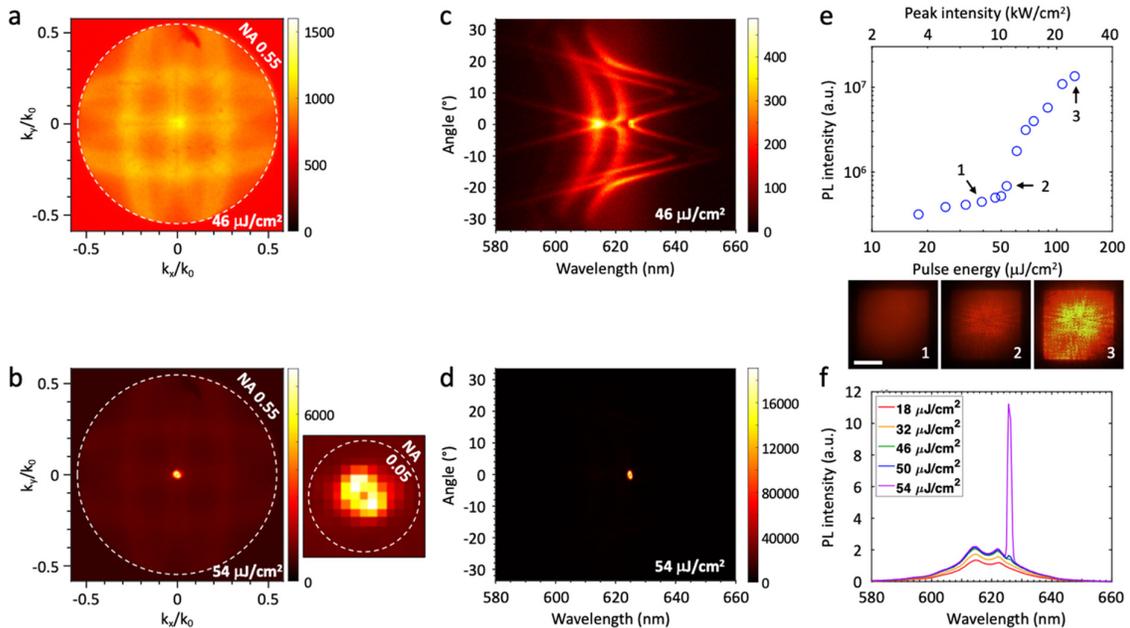

**Figure 4.** (a, b) Back focal plane image (radiation pattern) (a) below and (b) above the lasing threshold for a CQD-coated $TiO_2$ array with a cylinder diameter of 340 nm, at the pump fluence
9

indicated. The structure supports a BIC arising from a 2nd-order slab waveguide mode. The sample is excited by a pulsed laser (pulse width ~ 5 ns, repetition rate ~1250 Hz) at $\lambda = 355$ nm, and the PL is collected without a polarizer. The range of collection is limited by the numerical aperture (NA) of the objective. $k_x/k_o$ and $k_y/k_o$ are normalized in-plane wavevectors of the outcoupled light, indicating the direction of emission. The side panel in (b) is a zoomed-in image showing a donut-shaped radiation pattern in the normal direction. (c, d) Angle-resolved PL spectra (c) below and (d) above the lasing threshold at the pump fluence indicated, corresponding to (a) and (b) along $k_x = 0$, respectively. (e) Dependence of the CQD PL intensity on the pump fluence, plotted in a log-log scale, showing a lasing threshold of ~ 54 µJ/cm$^2$, equivalent to a peak intensity of ~ 11 kW/cm$^2$. The lower panel presents the fluorescence images of the array corresponding to the numbered data points; scale bar: 20 µm. (f) PL spectra (integrated across all angles) at various pump fluences showing the onset of lasing.

To characterize lasing, we photoexcite the sample with a pulsed laser at $\lambda = 355$ nm (pulse width ~ 5 ns, repetition rate ~ 1250 Hz). We perform back focal plane (BFP) imaging to capture the PL intensity as a function of the emission angle, to a maximum angle of ~ 33° limited by the numerical aperture of the collection objective (NA = 0.55). Figure 4a shows the BFP image of the unpolarized PL from the CQD-coated array with a cylinder diameter of 340 nm, supporting the 2nd-order BIC. Here, we see the 1st- and 2nd-order diffraction bands coupled to the CQD PL, with the 1st-order diffraction (i.e. 2nd-order DFB) approaching the normal. As mentioned earlier, passing the BFP image through the slit of a spectrometer yields the angle- and wavelength-resolved PL spectrum. Figure 4c shows the PL analyzed spectrally along the y-axis of the BFP. At a pump fluence of 46 µJ/cm$^2$, we see some enhancement of the PL intensity at the 2nd-order BIC, but the PL coupled to other bands is of a similar order of magnitude in intensity.

As we increase the pump fluence to 54 µJ/cm$^2$, the BFP image (Figure 4b) and the PL spectrum (Figure 4d) change dramatically. The PL intensity at the BIC increases by about 40 times and lasing occurs at $\lambda$ = 626 nm. Closer examination of the BFP image above the lasing threshold (Figure 4b, side panel) reveals a donut-shaped bright spot in the normal direction; see the angular distribution in Figure S2 in the Supporting Information. Intuitively, the donut shape is a result of suppressed radiation in the strict normal direction for the BIC, but lasing radiation in the vicinity of the BIC, where a good balance exists between a high Q factor and an accessible radiation channel. Such a radiation pattern is characteristic of lasing through BICs based on dipoles oscillating normal to the lattice plane.[11,14,21,22,27,34] Indeed, as shown in Figure 2d, the waveguide-based BICs we observe here have electric fields similar to those of out-of-plane magnetic dipoles. Lastly, the light-in, light-out curve (Figure 4e) indicates a lasing threshold of ~ 54 µJ/cm$^2$, corresponding to a peak intensity of about 11 kW/cm$^2$ for 5 ns pulses.

On the other hand, we also observe a lasing threshold of ~ 150 µJ/cm$^2$ for the array with a cylinder diameter of 280 nm, via the 1$^{st}$-order BIC; see Figure S3 in the Supporting Information for details. We note from Figure 2c that the field antinodes for the 1$^{st}$-order BIC lie mainly inside the TiO$_2$ cylinder, while the field antinodes for the 2$^{nd}$-order BIC lie also in the CQD gain medium. As a result, the 1$^{st}$-order BIC has poorer spatial overlap with the CQDs, which leads to a higher lasing threshold.

The threshold of 11 kW/cm$^2$, observed from the 2$^{nd}$-order BIC, is comparable to that of the CW lasing of the same type of CQDs reported earlier.[3] In the previous work, a 200 × 200 µm magnesium fluoride (MgF$_2$, $n$ ~ 1.4) grating was used on a MgF$_2$ substrate cooled to about -20 °C.[3] Here, we use a smaller 50 × 50 µm TiO$_2$ grating on a quartz substrate and at room temperature, but photoexcited using 5 ns pulses at 1250 Hz instead of at CW. While it is unclear whether the previous CW study involved a diffractive band-edge mode or a BIC, in this work, we identify the lasing mode specifically as a BIC, based on a 2$^{nd}$-order slab waveguide mode. This type of mode possesses not only a giant Q factor but also good spatial overlap with the CQDs, allowing us to achieve low-threshold, room-temperature nanosecond lasing.

Although our study suggests that the high-Q BIC is more favourable for lasing than the diffraction-coupled band-edge mode, it is important to note that more work is necessary to ascertain if BICs would indeed dominate (i.e. lead to a lower lasing threshold) under all circumstances, particularly given that the spatial mode-gain overlap can be better for the diffractive band-edge mode (Figure 1a) compared to the BIC (Figure 1b). Provided the spectral mode-gain overlap is similar, the competition between the Q factor and the spatial overlap would be interesting to explore. We advocate for characterization of lasing modes in detail, including the dispersion diagram and radiation pattern, for $2^{nd}$-order DFB lasers in the future. Understanding the nature of the lasing cavity should be helpful not only for device optimization but also for applications. For instance, BIC lasers are promising as compact, on-chip sources of optical vortex beams,[14,35–37] with their far-field radiation carrying well-defined topological charges (i.e. polarization rotating about the beam axis; see Figure 2d). Vortex microlasers at optical frequencies offer potential benefits in optical communications and quantum information processing.

In conclusion, we report a low-threshold laser at room temperature, consisting of a ~ 300 nm-thick film of CQDs covering a square array of $TiO_2$ cylinders. Lasing is achieved from high-Q modes known as BICs, arising from $1^{st}$ and $2^{nd}$-order TE-polarized slab waveguide modes, excited in the CQD-$TiO_2$ structure and coupled to the periodicity of the cylinder array. The $2^{nd}$-order BIC, in particular, gives a low lasing threshold of approximately 11 kW/$cm^2$ under 5 ns-pulsed optical pumping, thanks to a high Q factor and good spatial mode-gain overlap. In addition, for what is conventionally termed a $2^{nd}$-order DFB mode, we differentiate between a BIC and a band-edge mode coupled to diffraction. We show that BICs are the modes responsible for lasing in the (near) normal direction in our system and encourage researchers in the field to collectively verify if BICs, with their extraordinarily high Q factors, are the dominant lasing modes in $2^{nd}$-order DFB lasers. Finally, we highlight the potential application of BIC lasers, with their exotic radiation patterns, as sources of optical vortex beams for communications and information processing.

## ASSOCIATED CONTENT

**Supporting Information**

Methods; integrated PL spectrum of a CQD film; angular distribution of the lasing radiation from the $2^{nd}$-order BIC; lasing characterization of the $1^{st}$-order BIC.

## AUTHOR INFORMATION

**Corresponding Authors**

*Email: arseniy_kuznetsov@imre.a-star.edu.sg

*Email: ted.sargent@utoronto.ca

**Author Contributions**

M.W. fabricated the sample, performed optical characterizations, and wrote the first draft of the manuscript. L.K.S. synthesized the colloidal quantum dots (CQDs). G.B. deposited the CQD film and helped with initial optical measurements. L.D. performed numerical simulations under the guidance of R.P.-D. R.P.S. provided technical guidance. E.H.S and A.I.K supervised the work. All authors contributed to writing the manuscript.

**Notes**

The authors declare no competing financial interest.

ACKNOWLEDGEMENTS

This work was supported by the A*STAR SERC Pharos programme (grant number 152 73 00025; Singapore), AME Programmatic Grant No. A18A7b0058 (Singapore), National Research Foundation of Singapore under Grant No. NRF-NRFI2017-01, the Ontario Research Fund-Research Excellence Program, and the Natural Sciences and Engineering Research Council (NSERC) of Canada. In addition, M.W. acknowledges generous support from the Banting Postdoctoral Fellowship. The authors are grateful to Dr. Ha Son Tung and Joao Martins de Pina for assisting with optical characterizations, and Dr. Sjoerd Hoogland and Prof. Oleksandr Voznyy for helpful discussions.

**TOC Figure**

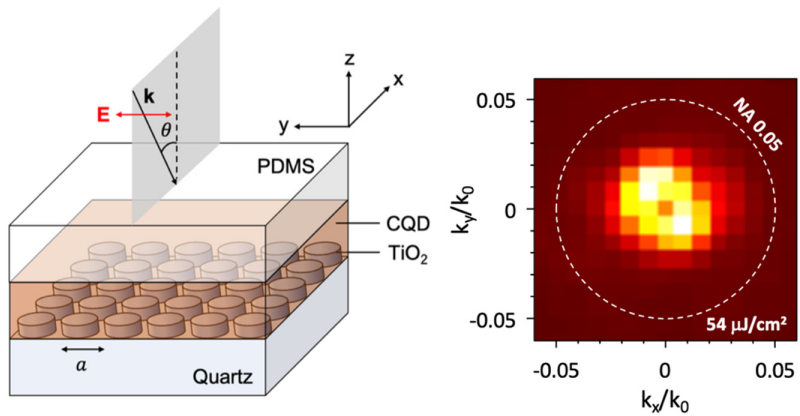

Microlaser. *arXiv:2107.05239* **2021**.

(28) Kuznetsov, A. I.; Miroshnichenko, A. E.; Brongersma, M. L.; Kivshar, Y. S.; Luk'yanchuk, B. Optically Resonant Dielectric Nanostructures. *Science* **2016**, *354*, aag2472.

(29) Kruk, S.; Kivshar, Y. Functional Meta-Optics and Nanophotonics Governed by Mie Resonances. *ACS Photonics* **2017**, *4*, 2638–2649.

(30) Staude, I.; Pertsch, T.; Kivshar, Y. S. All-Dielectric Resonant Meta-Optics Lightens Up. *ACS Photonics* **2019**, *6*, 802–814.

(31) Murai, S.; Abujetas, D. R.; Castellanos, G. W.; Sánchez-Gil, J. A.; Zhang, F.; Rivas, J. G. Bound States in the Continuum in the Visible Emerging from Out-of-Plane Magnetic Dipoles. *ACS Photonics* **2020**, *7*, 2204–2210.

(32) Azzam, S. I.; Chaudhuri, K.; Lagutchev, A.; Jacob, Z.; Kim, Y. L.; Shalaev, V. M.; Boltasseva, A.; Kildishev, A. V. Single and Multi-Mode Directional Lasing from Arrays of Dielectric Nanoresonators. *Laser Photonics Rev.* **2021**, *15*, 1–8.

(33) Sadrieva, Z. F.; Sinev, I. S.; Koshelev, K. L.; Samusev, A.; Iorsh, I. V.; Takayama, O.; Malureanu, R.; Bogdanov, A. A.; Lavrinenko, A. V. Transition from Optical Bound States in the Continuum to Leaky Resonances: Role of Substrate and Roughness. *ACS Photonics* **2017**, *4*, 723–727.

(34) Hakala, T. K.; Rekola, H. T.; Väkeväinen, A. I.; Martikainen, J. P.; Nečada, M.; Moilanen, A. J.; Törmä, P. Lasing in Dark and Bright Modes of a Finite-Sized Plasmonic Lattice. *Nat. Commun.* **2017**, *8*, 13687.

(35) Zhen, B.; Hsu, C. W.; Lu, L.; Stone, A. D.; Soljačić, M. Topological Nature of Optical Bound States in the Continuum. *Phys. Rev. Lett.* **2014**, *113*, 257401.

(36) Doeleman, H. M.; Monticone, F.; Den Hollander, W.; Alù, A.; Koenderink, A. F. Experimental Observation of a Polarization Vortex at an Optical Bound State in the
18

Supporting Information

# Bound State in the Continuum in Slab Waveguide Enables Low-Threshold Quantum-Dot Lasing

Mengfei Wu[1,2], Lu Ding[1], Randy P. Sabatini[2], Laxmi Kishore Sagar[2], Golam Bappi[2], Ramón Paniagua-Domínguez[1], Edward H. Sargent[2,*], Arseniy I. Kuznetsov[1,*]

[1] Institute of Materials Research and Engineering, A*STAR (Agency for Science, Technology and Research), Singapore 138634, Singapore

[2] Department of Electrical and Computer Engineering, University of Toronto, Toronto, Ontario M5S 3G4, Canada

*Corresponding authors: arseniy_kuznetsov@imre.a-star.edu.sg; ted.sargent@utoronto.ca

**METHODS**

**Synthesis of colloidal quantum dots (CQDs)**

**Chemicals and synthesis.** Cadmium oxide (> 99.99%), sulfur powder (S, > 99.5%), selenium powder (Se, > 99.99%), oleylamine (> 98% primary amine), octadecene (90%), oleic acid (90%), tri-butylphosphine (97%), trioctylphosphine oxide (99%), octadecylphosphonic acid (97%), 1-octanethiol (> 98.5%), thionyl chloride (SOCl2), toluene (anhydrous, 99.8%), hexane (anhydrous, 95%), acetone (99.5%) and acetonitrile (anhydrous, 99.8%) were purchased from Sigma-Aldrich and used without further purification. Trioctylphosphine (90%) was purchased from Alfa Aeaser.

**CdSe core synthesis.** CdSe CQDs were synthesized using existing literature protocol.[1,2] An amount of 120 mg CdO, 12 g TOPO and 0.56 g ODPA were mixed in a 100-mL three-neck flask, the mixture was heated to 120 °C for 30 min under vacuum, and the temperature was then brought to 320 °C and kept at that temperature for 2 h under nitrogen. A measure of 2 mL of TOP was injected into the flask and the temperature was further raised to 375 °C. The injected selenium precursor consisted of 2mL selenium in TOP solution at a concentration of 60 mg/mL. CQDs that exhibited an excitonic peak at 590 nm were produced as a result of 3-min growth time. The reaction was terminated by removing the heating mantle and adding acetone. The CQDs were purified by centrifuging at 6000 rpm for 3 min. Colorless supernatant was discarded and the resultant nanoparticles were redispersed in hexane for shell growth.

**Shell precursor synthesis.** Cd-oleate was prepared in a 250 mL flask by adding 0.98 g cadmium oxide and 40 ml oleic acid. The reaction mixture was evacuated under vacuum at 120 °C for 120 min to remove traces of water and other gases. Later, the reaction mixture was switched to nitrogen and temperature was increased to 250 °C to obtain Cd-oleate. TOP-S was prepared by adding 960 mg of sulfur in 16 mL trioctylphosphine under continuous stirring inside the glovebox.

**Biaxially strained asymmetric core/shell QDs synthesis.** The synthesis was done by following previously published report.[3] Our experience suggested that facet selective epitaxy is better achieved by distilling out ODE and oleylamine to remove any impurities. The first asymmetric shell was grown as follows. By measuring the absorbance at peak exciton (590 nm) with 1 mm light path length cuvette, we quantified CdSe CQDs. A 5.8 mL CdSe CQD in hexane dispersion with an optical density of 1 at the exciton peak was added to a mixture of 42 mL octadecene and 6 ml oleylamine in a 500 mL flask, and pumped in vacuum at 100 °C to evaporate hexane; then the solution was heated to 300 °C and kept for 30 min. As-prepared 9 mL Cd-oleate was diluted in 15 mL octadecene and 3 mL TOPS in 21 mL octadecene as a sulfur precursor, respectively. Cd-oleate and TOPS solutions were injected simultaneously and continuously at a rate of 6 mL h$^{-1}$. The second uniform shell was grown as follows: 4 mL Cd-oleate diluted in 20 mL octadecene and 427 μL 1-octanethiol diluted in 23.6 mL octadecene were continuously injected at a speed of 12 mL h$^{-1}$ to grow the second shell. The reaction temperature was elevated to 310 °C before injection. After a 13 mL injection of Cd-oleate into octadecene solution, 5 mL oleylamine was injected into the solution to improve the dispersibility of the CQDs.

**Biaxially strained asymmetric core/shell QDs purification.** When the injection was complete, the final reaction mixture was naturally cooled to around 50 °C and transferred into four 50 mL plastic centrifuge tubes; no anti-solvent was added and the precipitation was collected after 3 min centrifugation at a speed of 6,000 rpm. A total of 20 mL hexane was added into the centrifuge tubes to disperse the CQDs, and acetone was added dropwise until the CQDs started to aggregate. The precipitation was collected again by 3 min centrifugation at a speed of 6,000 rpm.; this dispersing and precipitation process was repeated three times to completely remove smaller CdS CQDs. **Note:** this purification process is critical for obtaining low-ASE-threshold and high-PLQY CQDs. The final CQDs were re-dispersed in octane with the first exciton peak absorbance in 1 mm light path length fixed as 0.25.

To achieve chloride ligand exchange,[1,3] 500 μL of the above CQDs dispersion (first peak exciton OD = 0.25) was vacuum dried and then dispersed in 1 mL toluene solution, 1.25 mL tributylphosphine, followed by 1 mL $SOCl_2$ in toluene solution (volume ratio of 20 μL $SOCl_2$:1 ml toluene) was added into the CQDs in toluene dispersion inside the glovebox. The CQDs precipitated immediately and the resulting dispersion was transferred out of the glovebox and subsequently ultrasonicated for 1 min. After the exchange, anhydrous hexane was added to precipitate the CQDs completely before centrifugation at 6,000 rpm. The CQDs were purified with three cycles of adding anhydrous acetone to disperse the CQDs and adding hexane to precipitate the CQDs dispersion. The chloride-ligand-terminated CQDs were finally dispersed in 750 μL anhydrous acetonitrile for the fabrication of the devices.

**Fabrication of $TiO_2$ nanostructures**

Fused silica (quartz) substrates were cleaned by ultra-sonication in deionized (DI) water, acetone, and isopropanol. 120 nm-thick $TiO_2$ was deposited by ion-assisted sputtering (Oxford Optofab3000), followed by 30 nm-thick chromium (Cr) by electron-beam (e-beam) evaporation (Angstrom EvoVac). To make $TiO_2$ cylinder arrays, the substrate was first heated at 140 °C for 1 min to improve adhesion of the e-beam resist on Cr. A negative e-beam resist, hydrogen silsesquioxane (HSQ, Dow Corning XR-1541-006) was spin coated at 5000 rpm for 60 s, followed by baking at 140 °C for 3 min. This led to an HSQ film thickness of about 120 nm. We then performed e-beam lithography (Elionix ELS-7000) and developed HSQ by immersing the sample in a salty developer[4] (1 wt.% NaOH and 4 wt.% NaCl in DI water) for 4 min, followed by a generous rinse in DI water. Using inductively coupled plasma reactive ion etching (ICP-RIE, Oxford PlasmaPro 100 Cobra), the Cr layer was etched, transferring the HSQ pattern, with a mixture of $Cl_2$ and $O_2$ gases ($Cl_2$–19 sccm, $O_2$–2 sccm, 10 mTorr). Then, using Cr as a hard mask (since HSQ would be etched together with $TiO_2$), the $TiO_2$ layer was etched with $CHF_3$ gas (25 sccm, 25 mTorr). Finally, Cr was removed by immersing the sample in liquid Cr etchant (Sigma Aldrich) for 4 min with gentle agitation. The sample was rinsed by DI water and IPA and blow-dried with $N_2$.

**Deposition of CQD film**

The quartz substrate patterned with $TiO_2$ cylinder arrays was cleaned with acetone and isopropanol, blow-dried with $N_2$, and treated with $O_2$ plasma for 5 min. The CQD solution was highly concentrated (~ 50 mg/mL) in octane and was filtered by centrifuging at 13,000 rpm for 2 min before spin coating. 50 µL of the CQD supernatant was spun in air at 1500 rpm for 60 s. The sample was then covered with a piece of polydimethylsiloxane (PDMS, ~ 0.1 mm thick) and stored in a glovebox before optical characterization under ambient conditions.

**Optical characterization**

The angle-resolved reflectance and photoluminescence (PL) spectra were measured by back focal plane (BFP) spectroscopy. The sample was positioned at the image plane of an inverted optical microscope (Nikon Ti-U) with a long-working-distance microscope objective (50×, NA 0.55). For reflectance measurements, the incident light was from a halogen lamp through the 50× microscope objective and a linear polarizer. For PL measurements, the sample was excited with a continuous-wave laser at $\lambda = 488$ nm through the same 50× microscope objective. For lasing measurements, the sample was excited by a pulsed laser at $\lambda = 355$ nm (pulse width ~ 5 ns, repetition rate ~ 1250 Hz) through a top 10× microscope objective, with a pump spot diameter of ~ 60 μm. The reflectance or PL signal from the sample was collected through the bottom 50× microscope objective, a linear polarizer (where applicable), and appropriate filters (where applicable), followed by a series of lenses that imaged the back focal plane of the 50× objective onto the entrance slit of a spectrometer (Andor SR-303i). The slit was 100 μm in width and was aligned with the x axis of the sample, i.e. collecting the emission in the x-z plane. A grating with 150 lines/mm, blazed at 500 nm dispersed the light after the slit, resulting in an angle- and wavelength-resolved reflectance or PL map, captured on a 2D EMCCD camera (Andor Newton 971). The spectral resolution was approximately 2 nm.[8]

**Numerical simulation**

Simulations were carried out using the Finite Element Method (COMSOL Multiphysics v5.5). For that, a single unit cell was simulated and Bloch boundary conditions applied in the lateral directions, to mimic an infinitely extended system. This system was excited, and the reflection and transmission recorded, using a set of periodic ports (one for each diffraction order supported by the system in the studied angular and wavelength span). The material parameters of $TiO_2$ and CQDs were taken from ellipsometry measurements, which closely reproduced those reported in literatures.[5] The external medium was considered homogeneous $SiO_2$ with refractive index taken from COMSOL library (Ghosh1999), thus representing the quartz substrate and the index-matching PDMS cover used in the experiments. The photonic modes were identified by field distributions. The quality factor of the resonance was obtain by fitting the reflectance to the Fano formula:[6,7]

$$R = A\frac{\left(2\frac{\omega - \omega_0}{\Gamma} + q\right)^2}{1 + \left(2\frac{\omega - \omega_0}{\Gamma}\right)^2} + B$$

where A and B are constants, q is the Fano asymmetry factor, $\omega_0$ and $\Gamma$ are resonant frequency and linewidth at half-maximum, respectively. Quality factor is obtained from $\omega_0/\Gamma$.

16404.

(8) https://andor.oxinst.com/resolution-calculator

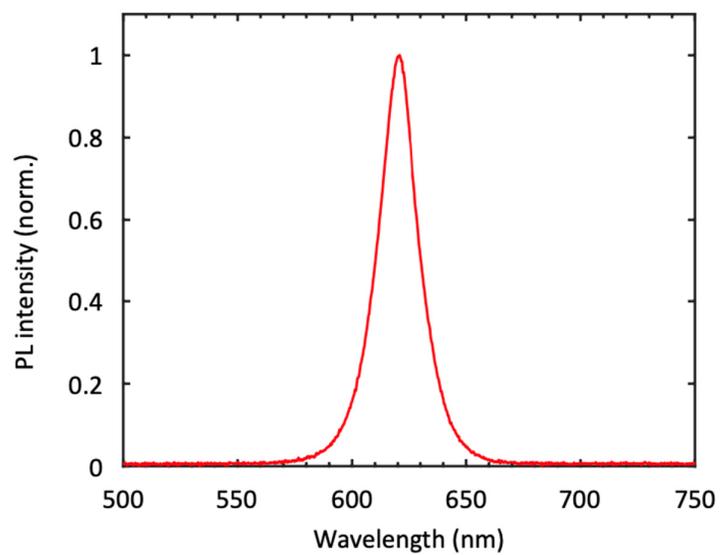

**Figure S1.** Photoluminescence (PL) spectra of CdSe/CdS core-shell colloidal quantum dots (CQDs) spin-cast as a solid film. PL peaks at $\lambda = 620$ nm with a full width at half maximum (FWHM) of 20 nm when the film is excited at $\lambda = 488$ nm.

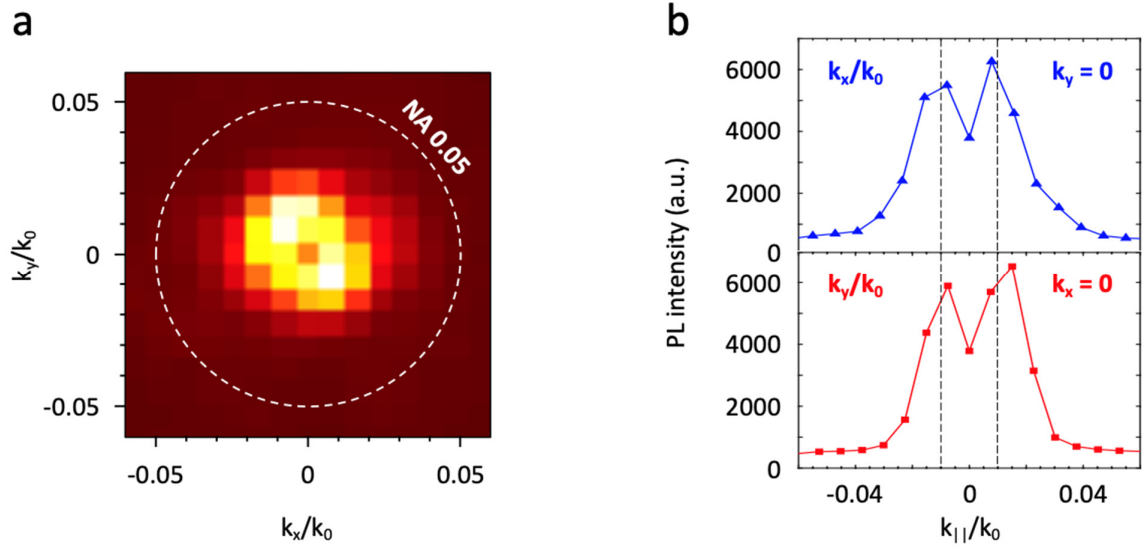

**Figure S2.** (a) Zoomed-in image of the far-field radiation pattern during lasing for a CQD-coated TiO$_2$ array with a cylinder diameter of 340 nm, corresponding to Figure 4b. The donut shape in the normal direction confirms that lasing arises from a bound state in the continuum (BIC). The direction of radiation is denoted by the ratio between the projected in-plane momentum k$_\parallel$ (i.e. k$_x$, k$_y$) and free-space momentum k$_0$. (b) Upper and lower panels present the line scans along the x and y axes of (a), respectively, showing the angle-resolved lasing intensity. Peaks in lasing intensity are observed at k$_\parallel$/k$_0$ ~ ±0.01 (dashed lines), corresponding to an angle of ~ 0.6° into the free space, or ~ 0.3° in the CQD layer with an index of 1.9.

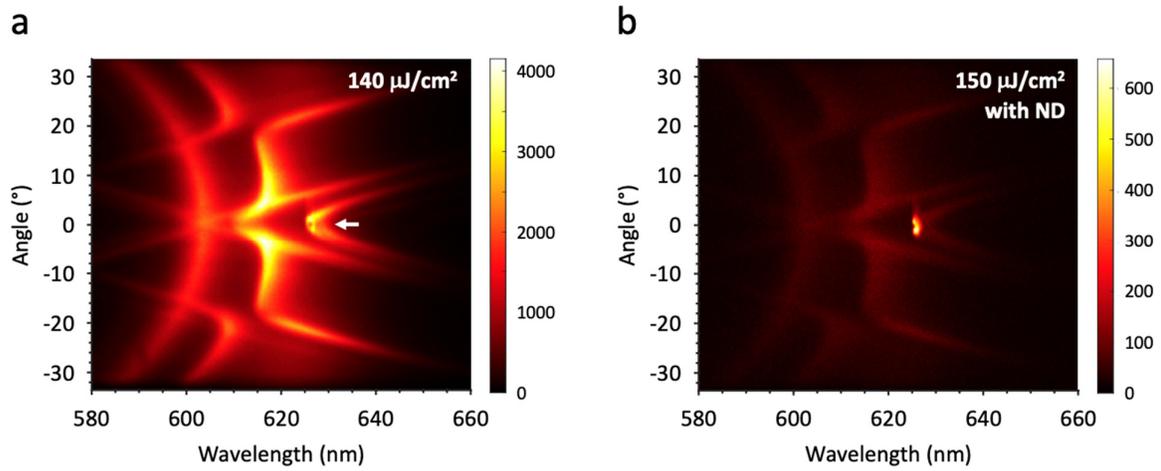

**Figure S3.** (a, b) Angle-resolved PL spectra (a) below and (b) around the lasing threshold for a CQD-coated $TiO_2$ array with a cylinder diameter of 280 nm, at the pump fluence indicated. The structure supports a bound state in the continuum (BIC, pointed by an arrow) arising from the 1st-order slab waveguide mode. The sample is excited by a pulsed laser at $\lambda = 355$ nm (pulse width ~ 5 ns, repetition rate ~ 1250 Hz), and the PL is collected without a polarizer. Lasing occurs via the 1st-order BIC at a pump fluence of ~ 150 µJ/cm². The spectrum in (b) is taken with a neutral density (ND) filter (optical density ~ 2).